# Instruments without optics: an integrated photonic spectrograph

J. Bland-Hawthorn[a], A. Horton

Anglo-Australian Observatory, 167 Vimiera Rd, Eastwood, NSW 2122, Australia

## ABSTRACT

In recent years, a great deal of emphasis has been placed on achieving the diffraction limit with large aperture telescopes. For a well matched focal-plane instrument, the diffraction limit provides the highest possible angular resolution and sensitivity per pixel. But it offers another key advantage as we now show. Conventionally, as the telescope aperture $D$ grows, the instrument size grows in proportion to $D$, and the cost increases as $D^2$ or faster. However, an instrument that operates at the diffraction limit can break the trend of spiralling costs. In traditional instruments, the light must pass through a succession of large lenses, mirrors and gratings, making it difficult to conserve the integrity of such a small psf. An alternative approach, as we now show, is to couple the diffraction-limited beam directly into an integrated photonic spectrograph operating in low-order modes.

## 1. INTRODUCTION

It is only in recent years that infrared telescopes have come close to achieving the diffraction limit, most notably at wavelengths longer than 2 microns. The Earth's turbulent atmosphere smears out information from space and this must be corrected for with real-time compensation using adaptive optics (AO). The diffraction limit defines the smallest *angular* detail on the sky that a telescope can resolve (1.22 $\lambda/D$ radians, where $D$ is the telescope diameter, and $\lambda$ is the wavelength of light). In the diffraction limit, the sky background is *minimized* within a detector pixel (due to the extreme magnification), but not completely blocked. But the diffraction limit offers another advantage that has yet to be realized in practice.

Traditionally, instruments are designed to match the natural 'seeing' of the site. This means that as the telescope aperture $D$ grows, the size of the instrument grows in proportion to $D$, and the cost of the instrument increases as $D^2$ or faster. Scaling from existing concepts, a single spectrograph for an ELT will cost $50-100M, comparable to the cost of building today's largest telescopes (Russell et al 2004). Given that the astronomical community is now moving towards extremely large telescopes (ELT), the dependence of cost on $D$ poses a major problem to realizing practical ELT instruments. We now propose a revolutionary approach to astronomical instrumentation that builds on advances within the field of photonics.

When a distant star is imaged at the telescope focus, its *physical* spot size (in μm) at the focal plane is independent of $D$, in contrast to its *angular* size on the sky (=1.22 $\lambda/D$). Indeed, the physical size of the imaged star is given by $P = 1.22 \lambda F$ where $F$ is the focal ratio of the telescope (i.e. focal length divided by $D$). For a typical optical/infrared system ($F$=5) operating in the infrared ($\lambda$=1.5 microns), the spot size is about 10μm. This tiny spot size is a challenge to astronomical instruments operating at the diffraction limit. Contemporary instruments are hard pushed to conserve the integrity of a spot size of ~100μm as the light passes through a succession of large lenses, mirrors and gratings. A spot size of only 10μm requires that the optical surface shapes be maintained to much higher tolerances, which is extremely difficult and expensive to do.

Recent developments in photonics provide an entirely new approach to solving this problem. The light can be manipulated over much smaller baselines than in conventional spectrographs. The light is launched into an integrated photonic circuit, being dispersed over a region of order ~1cm rather than the hundreds of cms associated with conventional spectrographs. The resolving power of any interference device can be expressed as R = $m\eta$ where $m$ is the diffractive order of interference, and $\eta$ is the disperser finesse (basically the number of combining beams; see Bland-Hawthorn & Cecil 1996), equivalent to the number of independent spectral elements in a single order. A useful device we are exploring has R ~ 1000 and large grating facets used at moderately high order ($m > 10$), quite unlike the parameters of a conventional astronomical grating.

---

[a] Email jbh@aao.gov.au; phone +61 2 9372 4851; fax +61 2 9372 4880

## 2. THE PHOTONICS REVOLUTION

The emergence of *astrophotonics* – a field that lies at the interface of astronomy and photonics – is timely as it has the potential to solve a number of outstanding problems in astronomical instrumentation (Bland-Hawthorn, Englund & Edvell 2004; Haynes et al 2004; Leon-Saval et al 2005; Corbett & Allington-Smith 2005).

We now introduce a new concept to astronomy, the integrated photonic spectrograph, that has the potential to solve the problem of spiralling costs in building future astronomical instruments. In certain respects, the concept of an integrated photonic spectrograph already exists within photonics and telecomm research groups, although in order to achieve a device that is suitable for astronomy, a significant amount of research and development is required.

There are a number of devices that show promise for use in astronomy. However, we limit our discussion to just two photonic devices that exemplify critical functions relevant to an integrated photonic device: the photonic array waveguide grating (PAWG) and the photonic echelle grating (PEG). We are currently engaged in development work on both devices with bandwidth foundries that specialize in integrating photonic functions on a single chip.

The early fibre optic networks were based on data transfer at a single laser frequency in single-mode fibres (with multimode fibres finding occasional use in the last stages of the network). By 1995, it was clear that the projected need for much higher data rates would require dense wavelength division multiplexing (DWDM), i.e. independent data streams carried by many contiguous wavelength channels. The fact that DWDM is widely used today for long haul communication is a significant achievement when one recalls the dispersive properties (i.e. different wavelengths propagating at different speeds) of optical fibres.

A key requirement of DWDM is the ability to disperse the multiband signal within a specific input fibre or a set of input fibres into separate output channels and the ability to switch signals between input and output channels. This has given rise to a remarkable device called the photonic array waveguide grating, also known as the optical phased array (phasor), phase-array waveguide grating, or the waveguide grating router. The first such device was invented by M. Smit (1991) in the Netherlands. These are now commercially available in various incarnations. PAWGs were actually preceded by echelle gratings and came into being as a result of manufacturing difficulties with PEGs. But in recent years, there have been substantial advances in making PEGs ("gratings on a chip") as we now show.

We describe both devices in turn, and attempt a comparison of their relative merits for astronomy. The PEG devices win out in the final analysis, but the PAWGs incorporate critical functions that could be of great benefit when used in conjunction with PEGs.

## 3. PHOTONIC ARRAY WAVEGUIDE GRATINGS

PAWGs are remarkable, both in their complexity and in their operation (see Figs. 1 and 2). As far as we can determine, there has never been an approachable discussion of array waveguide gratings which may reflect the fact that these critical components are still evolving to the present day. Different parts of the evolving story can be found across a dispersed literature in optics, photonics, quantum electronics and communications journals. The two most basic configurations are called "symmetric" (Smit 1991) and "anti-symmetric" (Adar et al 1993) AWGs.

The AWG comprises input and output fibre ports, a multiplexor (mux) feeding a parallel array of closely spaced, single mode waveguides that in turn feed a demultiplexor (demux). The input[b] and output fibre ports are single-moded for the most part, although multimode operation has also been investigated. A commercially packaged device is an integrated circuit of order several centimeters in size, incorporating both mux and demux at either end of a waveguide array, with pigtails at both ends (input/output ports) to allow communication with the outside world.

Both the mux and demux comprise a sophisticated planar optic device called a "star coupler" (Somekh et al 1973). The star coupler is a slab waveguide with precisely manufactured concave edges defined along a Rowland circle, for reasons we return to later. The concave boundaries have a specific focal length $f_s$ and serve to focus the light onto the exit ports. Within the mux, the input port is bonded to one concave edge, and the array waveguide is bonded to the other. Similarly, within the demux, the array waveguide is bonded to one concave edge, and the output ports are bonded to the other. The slab waveguide is often referred to as the "free propagation zone" (FPZ) since the light is able to propagate freely; the constructive interference arises from the careful positioning of the input/output ports with respect to the phased waveguide array.

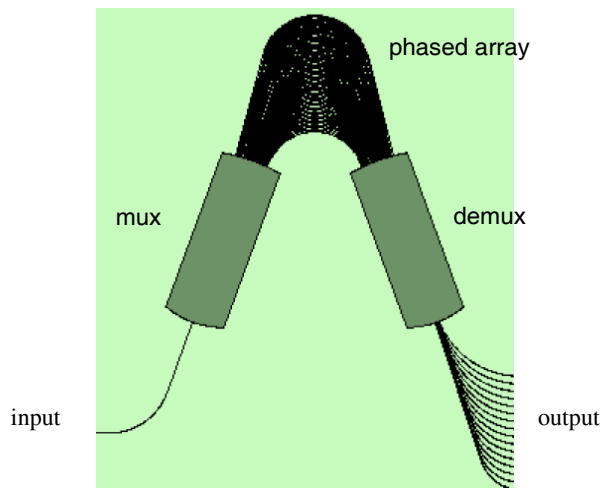

*Fig. 1.* Symmetric 1×N array waveguide grating. A single input fibre feeds light into the first slab waveguide (mux) that in turn directs the light onto an array of parallel waveguides. The parallel channels define a phased array and thus behave like a grating. On output, the grating produces interference in the second slab waveguide (demux). The system is designed so that different spectral channels are carefully imaged onto distinct output fibres.

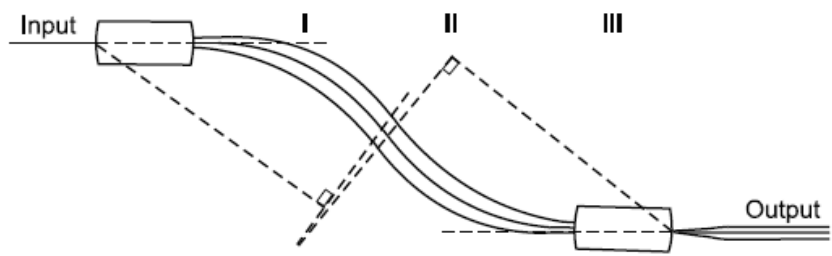

*Fig. 2.* Anti-symmetric 1×N array waveguide grating (KMBD[2]). The phased array is in 3 sections: the first section inserts an extra 60μm of path length but this is compensated for in the third section such that the overall pathlength difference between branches is determined solely by the straight middle section (the dashed lines are construction lines used in a design software package).

The input waveguide transports a range of wavelengths into the first star coupler (mux). The numerical aperture of the input fibre spreads out the light within the mux, but the light is directed onto the waveguide array. The number of channels in the waveguide array is limited by the width of the illumination pattern. If the numerical aperture of the input waveguide ($NA_{in}$) is matched to the NA of the mux ($NA_s$), this can help to minimize losses at the fibre-chip boundary.[c]

---
[b] Here we restrict ourselves to the specific case of a single input port.

Kok et al (2003; hereafter $KMBD^2$) present a multimode AWG that we use for illustration purposes here. For their device, the input fibre is a GRIN MMF with $NA_{in} = 0.2$ and core size of 40 μm; to ensure good coupling, the slab has a matched core size ($NA_s = 0.285$) leading to fibre-chip losses of about -1.3 dB after careful optimization. For a slab core of thickness $h_s$ and propagating wavelength $\lambda_s$ ($=\lambda_o/n_s$), it can be shown that the number of propagating modes is given by (McMullin et al 2002)

$$P = 1 + 2.NA_s.h_s/\lambda_s$$

or about ~20 modes in the case of the Kok device.

From geometrical optics[d], the theoretical width of the far-field illumination pattern is ~ $2 f_s NA_{mux} / n_s = 1770.5$ μm. $KMBD^2$ use this fact to design and manufacture the phased array. The slab waveguide incorporates a 40-μm thick polymer core between cladding layers made from a copolymer of HEMA and Styrene. These layers sit on a glass wafer (substrate) with a thickness of roughly 1 mm. The polymer core uses SU-8, a material that is highly transmissive at near-infrared wavelengths and responds well to UV photo-etching. The phased array of waveguides is usually realized with shallow ridges etched into the slab core. $KMBD^2$ photo-etch 127 parallel waveguides into the core material, each 10 μm wide, 40 μm high, and separated by 4μm. The incoming light then propagates through the mux before coupling into the discrete parallel waveguides. Some light (~30% or -1.5dB) is inevitably lost at this interface although this loss can be reduced to about -0.15 dB with more sophisticated designs, e.g. vertically tapered AWGs (Sugita et al 2000; Okamoto 2006, Fig. 9.65).

The number of parallel waveguides is usually much larger than required by the measured finesse in these devices. As we show below, commercial AWGs conform to industrial standards, ease of mass production and general applicability. They are not highly optimized devices in the sense of astronomical gratings. A typical commercial device has five times as many array channels when compared to the number of output channels. This is partly to ensure minimal cross talk between the output channels, a requirement that can be softened considerably in astronomical devices. Here, the typical cross-talk suppression between distinct spectral channels is -4 dB rather than the >30 dB demanded by telecommunications.

**Grating.** The phased array of waveguides is critical to the operation of the AWG. It behaves as a grating in that path length differences are imposed by the different branches of the array. In the symmetric layout (Fig. 1), where the channels are defined by a series of parallel arcs, the inner channel has the shortest pathlength. The next channel above it is longer by an optical path length $\delta L$, and the third channel is longer by $2.\delta L$ and so on till we reach the outermost waveguide. The path length difference between adjacent branches equals an integer multiple $m$ of the vacuum central wavelength ($\lambda_o$) defined by the middle channel on output. We can express this as

$$\delta L = m \lambda_o/n_a = m \lambda_a$$

where $n_a$ is the refractive index in the waveguide array and $\lambda_a$ is the propagating wavelength in the array (i.e. $\lambda_a=\lambda/n_a$).

A good illustration of what happens on output from a phased array is to consider the case of Young's classical 2-slit experiment in the limit of multiple slits (see Hecht 1990, Fig. 10.20). The far-field illumination is determined by two factors: (i) the width of a single channel (low-order envelope) and (ii) the total number of waveguides (high frequency structure) and the electric field distribution across them.

---

[c] These losses are typically of order $(NA_{in}/NA_s)^2$.
[d] The proper width must account for diffraction at the fibre-chip interface: this can narrow the far-field pattern even further. In fact, these authors measure a far-field width of only 1.2mm rather than the 1.8mm used in their design.

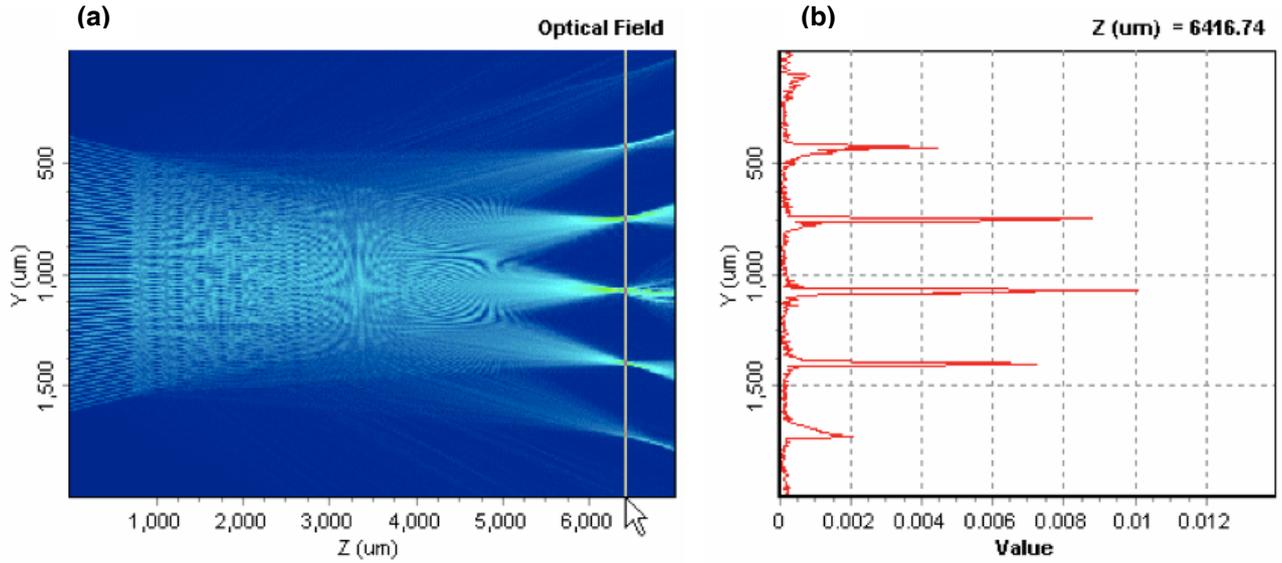

Fig. 3. Output from the APSS Apollo Photonics software package demonstrating (a) the near and far field pattern from an AWG at the design on-axis wavelength $\lambda_o$ = 1500nm at high order (*m*=120), (b) The y-axis intensity profile indicated by the arrow in (a) showing the different diffraction orders and the blaze envelope. The broad diffraction envelope arises from the far field response of a single array waveguide; the high frequency structure is due to the total number of waveguides within the array, i.e. one side lobe for each array waveguide.

To complicate matters slightly, the field in a given waveguide couples into the neighbouring waveguides as the field propagates forwards (Somekh et al 1973). This may explain why AWGs perform so well in terms of their sidelobe suppression characteristics compared to a normal diffraction grating. The field distribution in the array, which is typically a broad Gaussian across the branches (Fig. 4a), influences the wing structure in the output psf (Fig. 4b).

**Electric field distribution.** Let E=E(y,z) define the electric field of the waveguide array where z defines the axis of foreward propagation (see Fig. 3a). If the central waveguide (p=0) is excited at a distance $L_c$ before its exit point (boundary condition $E_o(0,-L_c)=\delta_o$), then the field distribution on exit arising from the $p^{th}$ waveguide is given by (Samekh et al 1973)

$$E_p = E_p(0,0) = i^p J_p(2\chi)$$

where $J_p$ is the $p^{th}$ order Bessel function of the 1$^{st}$ kind, and $\chi=\kappa L_c$ for which $\kappa$ is the coupling coefficient (~ 0.5 mm$^{-1}$). Marcatili (1969) provides a useful formalism for $\kappa$ in terms of the waveguide structure. To a good approximation, we can write for the overall field

$$E(y, z=0) \approx \sqrt[4]{(2/\pi w_o^2)} \sum_p i^p J_p(2\chi) \exp(-(y-p.d_a)^2/2.w_o^2) \qquad (1)$$

where the summation occurs over all 2M+1 waveguides[e]. Note that z=0 defines the spatial field along the Rowland circle (y direction) on exit from the phased array. Even though z=0 has a small amount of curvature in practice, the above approximation has been verified from full diffraction theory (e.g. Song & Park 2004). After imposing a suitable normalization (e.g. $\int |E(y,0)|^2 dy = 1$), we can plot this function for evolving z (forward direction through the demux) and identify the near-field Fresnel zone and the far-field Fraunhofer diffraction pattern (see Fig. 3). The output field profile results from a sum of Gaussian beams with identical (sigma) width $w_o$. Thus, for the central channel, the field distribution on input is reproduced at the central channel on output for the specific wavelength $\lambda_o$.

It is possible to recast equation (1) in terms of $\lambda$ rather than y at a fixed z value. This provides us with the detailed psf structure of a single resolution element on output (Figs. 3b and 4b). From Fourier optics, the intensity profile is a

---

[e] The waveguides are labelled p=0, ±1, ±2, …±M.

Fourier sum over all the channels of the phased array. For the $k^{th}$ receiver on output at an angle $\theta_k$ to the central axis, the intensity profile is $I(\lambda) = |E_k(\lambda)|^2$ where

$$E_k(\lambda) = \sum_j \text{sinc}^4[\pi/(2M+1)(j-M-1)] \exp[2\pi i\, jn_a(\delta L + d_a\theta_k)/\lambda] \quad (2)$$

The sinc function[f] assumes that the propagating electric field has a flat intensity distribution, but this can be Gaussian or somewhere between in practice (Fig. 4a). The exponential part provides the phase contribution. The summation in j (for which j=1 corresponds to the shortest waveguide) is over all 2M+1 waveguides in the phased array. An example of an AWG spectral psf is shown over a full free spectral range in Fig. 4b.

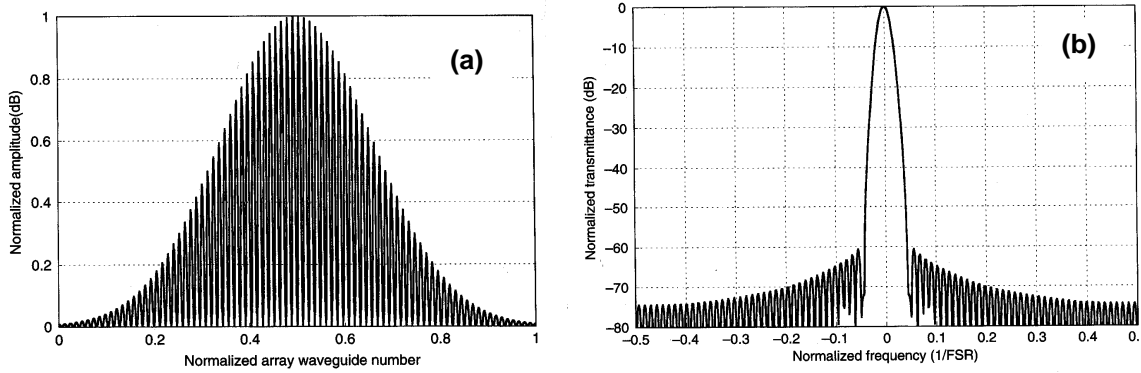

*Fig. 4.* (a) Electric field distribution across the waveguide array. (b) Electric field distribution at the focal surface for a single output resolution element.

**Spectral dispersion.** Now consider a wavelength that is offset by a small amount $\delta\lambda_o$ (in vacuo) from the central wavelength, i.e. within the free spectral range of the star coupler[g]. In this instance, $\delta L = m(\lambda_g + \delta\lambda_g)$ which causes a phase shift that increases linearly from the innermost to the outermost waveguide. This results in a tilted wavefront on output (see Fig. 5). The tilt angle $\theta$ with wavelength of the focal point in the image plane is given by

$$d\theta/d\lambda = m\, n_G / (n_a\, n_s\, d_a)$$

for which $n_s$ is the refractive index of the slab waveguide, $d_a$ is the physical separation of the array branches (Fig. 5), and $n_G$ is the *group* refractive index of the waveguide array, i.e.

$$n_G = n_a - \lambda_o(dn_a/d\lambda_o) = n_a - \lambda_a(dn_a/d\lambda_a).$$

The last part follows from the fact that the wavelength measured in the waveguide is $\lambda_a = \lambda_o/n_a$. Thus, it follows that

$$\delta L = m\, \lambda_a$$
$$d\theta/d\lambda_a = m\, n_G / (n_s\, d_a)$$

---

[f] The exceptionally high power index on the sinc factor reflects two stages of fibre coupling, i.e. input to array, and array to output (see Adar et al 1993).
[g] The star coupler is an interference device and therefore exhibits periodicity for different orders of interference. The free spectral range is defined as $\Delta\lambda \approx \lambda/(m+1)$ where $|\lambda-\lambda_o| < \Delta\lambda$.

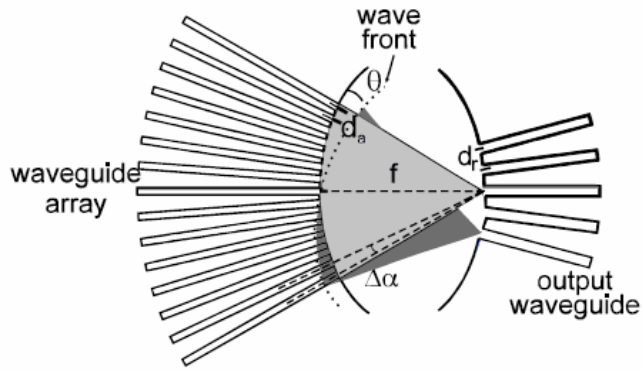

*Fig. 5.* An AWG is specified at a central (on-axis) wavelength $\lambda_o$. Any other wavelength within the operating free spectral range induces a wavefront tilt $\theta$ at the exit of the phased array. $\Delta\alpha$ is the divergence angle of the array branches and $f_s$ is the focal length of the slab waveguide. The pitches of the array waveguides and output waveguides are $d_a$ and $d_r$ respectively.

**Spectral resolution.** Separation (i.e. lateral displacement) of the wavelength channels is achieved by placing the output waveguides at specific locations in the demux image plane. We now seek a relation between the wavelength channel spacing $\delta\lambda$ and the physical separation between channels in the image plane, $\delta y$ ($= f_s \, \delta\theta$). The latter quantity depends on the array pitch distance $d_a$. We find that

$$\delta y = \lambda_a (dy/d\lambda_a) = \lambda_a f_s \, (d\theta/d\lambda_a) = \lambda_a \, n_G \, m \, f_s \, / \, (n_s \, d_a)$$

Similarly,

$$\delta\lambda = \delta\theta \, (d\theta/d\lambda)^{-1} = (\delta y / f_s) \, n_s \, n_a \, d_a \, / \, (m \, n_G)$$

The focal length of the slab wave guide (after setting $n_s = n_a$) is therefore

$$f_s \approx (\delta y / \delta\lambda_a) \, d_a \, / \, m \qquad (3)$$

or equivalently,

$$\delta y / \delta\lambda_a = m \, f_s \, / \, d_a \qquad (4)$$

From the above equations, we see that in order to achieve higher spectral resolution for a fixed output channel spacing $\delta y$, the focal lengths of the mux/demux must be increased.

For the device manufactured by KMBD[2], the AWG is specified for low order diffraction at a central wavelength of $\lambda_o$ = 850nm in vacuo. Both slab waveguides use $f_s$ = 4900 μm, and are operated in the 5$^{th}$ order ($m$ = 5). Thus, the free spectral range in vacuo is approximately $\Delta\lambda_o \approx \lambda_o/m+1 \approx 140$ nm.

Note from equation (1) that if we adopt the same waveguides that were used on input, the output channel configuration is largely specified. The authors set out to design a low order 1×3 AWG, and so to avoid inter-order confusion, $\delta\lambda_a = \Delta\lambda_a/3 = 140/(1.58×3) \sim 30$ nm. At the exit of the phased array, $d_a$ = 14 μm. Thus, we find $\delta y$ = 52.5μm (the quantity $d_r$ in Fig. 5) on exit from the slab waveguide. Adopting a more accurate estimate of $\Delta\lambda_o$, we find $\delta y$ = 55μm such that the 40μm output waveguides are separated by 15μm.

The measured spectral response profiles (in vacuo) are shown in Fig. 5 of KMBD[2]. All three bandpasses overlap at 3 dB (~45-50 nm FWHM) as expected. This may appear to produce a high degree of cross talk between the 3 output bands, unlike most commercially available devices working at higher diffraction order. But such a device is very effective when used as the pre-filter ahead of a parallel set of AWGs or PEGs working at much higher order. In this "tandem" configuration, each output is fed into a 1×N AWG, say, where N can be very large indeed (e.g. Takada et al 2002). Typically, the design of the second bank of AWGs suppresses cross talk between bands.

In Fig. 5 of KMBD[2], the peaks occur at $\lambda_A \approx 800$, $\lambda_B \approx 850$ and $\lambda_C \approx 900$ nm ($m$=5). We would expect the bands at lower and higher orders to follow the relation, $(m+1)\lambda_6 = m\lambda_5 = (m-1)\lambda_4$ such that $\lambda_6 \approx 700$nm and $\lambda_4 = 1050$nm. However, in practice, we must be wary of the dispersive properties of the waveguide material over such a large spectral range, i.e. the differential refractive index of the waveguides as a function of wavelength. In order to simplify the discussion, we have restricted all of the analysis to the propagation of light at the central wavelength $\lambda_o$ = 850 nm. The bands at lower and higher orders will have lower overall throughput compared to the middle order.

**Remaining issues.** There are a number of issues we have brushed over. First, the design of a phased array involves a number of extra parameters (see the discussion by Okamoto 2006). In the Kok device, the pathlength $\delta L$ between branches of the array is approximately 3µm which is much too small to help separate the waveguides. Instead, the Kok device exploits the anti-symmetric layout in Fig. 2. We note that the array is in three sections. The first section is extended with an extra pathlength of $\delta L_1$ = 60µm, the middle straight section uses the required phase difference of $\delta L_2 \approx$ 3µm, and the final section compensates for the first section with $\delta L_3$ = -60µm. The overall pathlength difference is $\delta L = \delta L_1 + \delta L_2 + \delta L_3 \approx$ 3µm. This approach to pathlength compensation is widely used in interferometers and birefringent filters at low order (e.g. Bland-Hawthorn et al 2001).

Secondly, the original intention of the Kok device was to develop an AWG that could operate with a multimode input. These authors demonstrate that a low-order device can tolerate a fair amount of modal dispersion without serious impact. This becomes more of an issue at higher diffraction order and higher spectral resolving powers. For any device, it is worth investigating how much power is transported in the high order modes. If these are found to be underfilled in multimode devices (e.g. Loke & McMullin 1990), it is possible to strip high order modes without serious loss of performance such that the device can be made to operate in multimode with minimal cross talk (e.g. McMullin et al 2002).

Thirdly, in a conventional AWG, the output ports sit along a Rowland circle in order to bypass the primary sources of optical aberration (see below). However, it is possible to design an AWG that produces a flat focal plane (e.g. Wang et al 2001; Lu et al 2003), an arrangement more suited to astronomical detectors. The free variables are the design of the slab waveguides and the differential pathlength differences between array branches. An alternative approach is to image the light onto a weakly curved infrared detector but these are not commercially available at the present time.

We have not addressed issues relating to TE-TM polarization and birefringence (see e.g. Spiekman et al 1996). Nowadays, polarization effects are largely understood and compensated for in AWG design (Okamoto 2006), and in the case of multimode devices are likely to be negligible (McMullin et al 2002).

We now discuss another important photonic device, the photonic echelle grating, before discussing the relative merits of these devices and their application to astronomy.

## 4. PHOTONIC ECHELLE GRATINGS

It has long been realized that photonic echelle gratings have major advantages over AWGs from a theoretical standpoint. In particular, the number of teeth in an echelle grating can be much larger than the number of branches in a phased array. As a result, the spectral finesse is much higher resulting in many more output channels. Another advantage is the relative compactness of PEG design, i.e. the path can be folded. Smaller devices means that the stringent requirements on material uniformity are easier to meet.

In the 1980s, industry struggled with the verticality and smoothness of deeply etched grating facets, inter alia, and as a result, AWGs came into being as an alternative approach to demultiplexing. But since that time, there has been great progress on all fronts such that PEGs are now undergoing a serious revival within the integrated photonics industry

(Delage et al 2004; Janz et al 2004). The devices can now be made in various materials, including $SiO_2$/Si, InGaAs/AlGaAs/GaAs and InGaAsP/InP. In addition to the facet problem, industry is now coming to grips with two other major challenges: (i) the polarization dependence of the grating efficiency, and (ii) compensation for birefringence. From the astronomer's perspective, the latter challenges are of secondary importance insofar as they do not compromise the overall throughput of the grating.

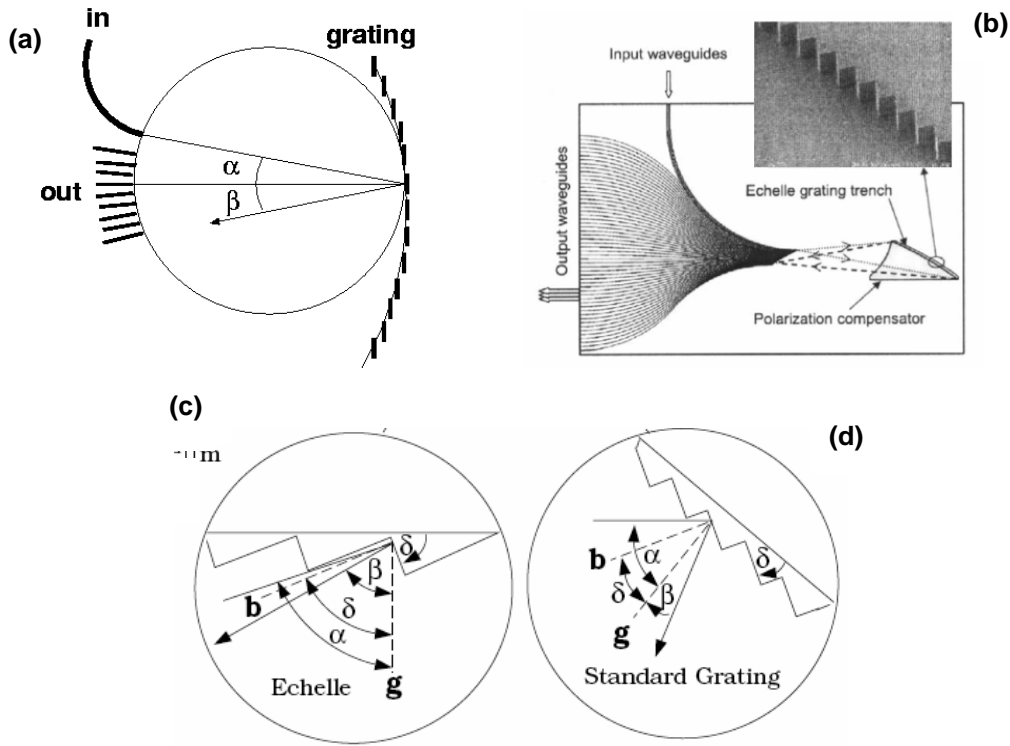

*Fig. 6.* (a) Schematic of a photonic echelle grating; the grating grooves are of order tens to hundreds of microns wide. (b) Photonic echelle grating (Janz et al 2004) designed to demultiplex the wavelength region 1530-1570nm into 48 output channels. The input spectrum enters from the top (see (a)) but is directed towards the grating grooves at a high angle of incidence compared to a conventional grating; compare (c) and (d) where **b** and **g** are the blaze and grating normals.

From an astronomical standpoint, PEGs are much closer to conventional astronomical gratings in terms of how the optical pathlength difference is introduced. However, in commercially available packages, the discrete wavelength intervals of interest are focused onto output waveguides (Fig. 6a). This ensures a very high degree of cross-talk suppression (-30 dB or more) between channels. In this respect, existing PEGs are overengineered and wasteful for astronomical use. We are content for the spectral dispersion to blend neighbouring resolution elements with cross talk approaching 25%, i.e. cross-talk suppression between channels approaching 75% (-4 dB suppression).

Fig. 6a illustrates a few×48 demux based on an echelle grating. Given that astronomers are mostly familiar with reflection gratings, we restrict our discussion to a basic overview. Thus we do not give a proper account at this stage of how the grating facets can be modified to improve the system performance. A highly readable account of how to simulate the grating electric field is given by McMullin et al (2002). A detailed discussion of the full Kirchhoff-Huygens diffraction theory is given by Marcuse (1982; 1991).

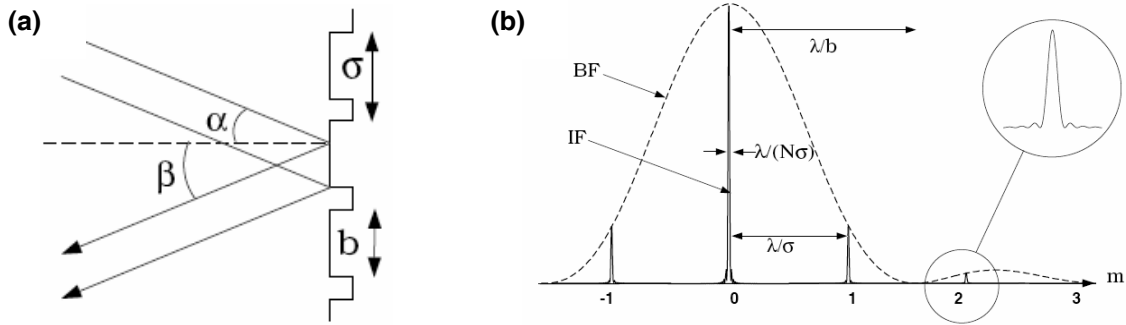

*Fig. 7.* (a) Grating angles and groove structure where $\sigma = d_a$. (b) Electric field distribution from a ruled diffraction grating for a monochromatic source in the focal plane. The influence of the groove structure (a) is also shown. The blaze function (BF) and interference function (IF) of the diffraction grating is indicated.

**Grating equation.** If a plane wave of wavelength $\lambda_a$ is incident at an angle $\alpha$ to the perpendicular of a periodic grating with groove spacing or pitch $d_a$, the outbound reflected beams at an angle $\beta$ interfere constructively according to

$$d_a (\sin \alpha + \sin \beta) = m \lambda_a \qquad (5)$$

The conventional formula for the grating dispersion $d\beta/d\lambda$ follows from differentiating this equation. However, in order to determine the positions of the output waveguides, a more useful formula expresses the wavelength dispersion with lateral displacement y measured in microns. At the Littrow angle (when light is reflected back on itself, i.e. $\alpha \approx \beta$), we can write

$$\delta y/\delta \lambda_o = 4 \rho \sin \alpha / \lambda_o (1-(\lambda_o/n_a)(dn_a/d\lambda_o)) \qquad (6)$$

where $\rho$ is the radius of the Rowland circle (Fig. 6). Once again, the "o" subscript refers to wavelengths in vacuo.

In an idealized astronomical grating, Fourier optics demonstrates that the field distribution in the focal plane is the Fourier power spectrum of the reflected signal across the beam. The grating acts as a filter that blocks all frequencies except those associated with its groove frequency or spatial frequency content within each groove. The number of grooves and maximum pathlength difference is set by the grating length L (Fig. 7a). The rectangle function that defines each grating groove produces a broad diffraction envelope (blaze function) that modulates the signal at each diffraction order, and whose width increases as the groove narrows. The Shah function that defines the extent of the grating and each groove produces a high-frequency diffraction envelope (or cluster – see Fig. 7b) at each diffraction order in much the same way as Fig. 4b (Gray 1992). The width of each cluster decreases as the grating length increases. Roughly speaking, the finesse of the curved grating is $\eta = L/d_a$ such that the maximum intensity of the secondary peak in a cluster is a fraction $\eta^{-2}$ of the primary peak.

The photonic echelle grating comes much closer to realizing the design finesse than an array waveguide grating which is one of its great strengths. But the PEG is quite unlike an astronomical grating in practice. The PEG comprises a series of mirror facets etched into the slab waveguide. These facets are much larger than what we conventionally associate with astronomical gratings where rulings are of order hundreds or thousands of lines per mm. The lithographic process allows many distinctly shaped facets (tens to hundreds of microns in size) to be positioned and oriented accurately so that a wide range of wavelengths are correctly focused to the appropriate output waveguide.

In PEGs, much like in conventional grating spectrographs, the blaze function BF can be used to modify the grating response. Time delays can be introduced across the grating to shift the peak of the diffraction envelope to an angle that corresponds to non-zero dispersive orders. The grating facets are tilted through an angle $\delta$ with respect to the grating normal. It can be shown that the blaze intensity profile has the form

$$BF(\lambda) = \text{sinc}^2[\pi d_a \cos \delta (\sin(\alpha-\delta)+\sin(\beta-\delta))/\lambda]$$

where the angles are defined in Figs. 6b and 6c.

**Aberrations.** Rowland (1883) first described a concave grating that was able to focus and diffract light simultaneously (see Beutler 1945 for a detailed analysis). Today, most commercially available PEGs and PAWGs are constructed on a Rowland circle in the sense that both the input and output channels lie along or close to this curve. But even the Rowland grating does not maintain phase coherence for widely separated points along the grating and the grating pitch must be modified to reduce aberrations (Poguntke & Soole 1993).

For astronomical purposes, it is highly desirable to achieve a flat focal plane such that the output focal points lie along a straight line. It is relatively straightforward to achieve a flat output facet by cleaving and polishing. But if the focal points lie along a circle, these will be progressively defocused and aberrated as we go off axis.

In the past decade, there has been a lot of discussion on how to flatten the focal surface while reducing aberration. One arrives at non-circular concave gratings defined by $2^{nd}$ or $4^{th}$ order differential equations. A highly readable account is given by McGreer (1995) although early attempts go back to Kastner & Wade (1978). The particular challenge is to control chromatic aberration given that the basic design adopts a central wavelength $\lambda_o$. Lightwaves at any other wavelength (within the design free spectral range) arriving from different parts of the grating must be kept in phase at the focal point.

**Grating size and miniaturization limits.** So how big is a photonic echelle grating for a given resolving power R? For a widely used grating angle of $\alpha=60^o$, the Littrow formula in equation (6) tells us

$$\rho \approx 0.3 \ \delta y \ R \approx 4.5 \ R \qquad \text{microns} \qquad (7)$$

where we have taken the output pitch to be about 15μm, i.e. the size of a pixel in the majority of high-performance infrared detectors[h]. For the flatfield broadband spectrograph design of McGreer (1995), the Rowland circle radius is about 20% larger (i.e. $\alpha=45^o$). The pitch is large compared to astronomical gratings because PEGs are mostly used at high order ($m > 10$; see McMullin et al 2002).

*We therefore arrive at a remarkable fact.* An astronomically useful photonic grating working at R~1000 needs only to be of order ~1 cm in size, far smaller than conventional diffraction gratings. Such devices are already commercially available for the telecomm market, except that these are overly sophisticated and lossy for astronomical purposes. Tens or hundreds of discrete wavelength channels are coupled into a matching number of output waveguides to ensure very low cross-talk between channels. But astronomers are content with continuously dispersed spectra, as long as the spectral psf is relatively well tapered (i.e. Gaussian[i]).

Ultimately, there are other considerations that determine the size of a PEG. First, there is the technology platform, e.g. wafer size and etching mask dimension. Secondly, as we have alluded to, the difficulty of achieving a high level of material uniformity may limit the size of the device. Thirdly, there are propagation losses which can be high for large devices or in devices exploiting high index contrast materials. In the best published grating devices, the insertion losses fibre to fibre are about -2 dB but this can be improved if we re-engineer them for astronomical use. The largest photonic circuits that can be manufactured at the present time are of order 10-15 cm; these glass devices have much better throughput characteristics than other materials. However, the cost per grating reaches its peak if the full wafer is used, since many more smaller devices can be printed onto a single wafer.

**How many output spectral channels?** At near infrared wavelengths, a highly desirable grating must deliver of order R resolution elements on output. Typically, the psf is Nyquist sampled and matched to the longest axis of a detector. A widely used detector today (e.g. Rockwell HAWAII series) has 1024 to 2048 pixels, such that ~1000 output channels can be detected simultaneously. Pixel sizes are in the range 10 to 20 μm and this is well matched to the lateral dispersion at the focal plane (see equation 6) with either moderate focal reduction or magnification.

---

[h] For Nyquist data sampling (e.g. Brault & White 1971), we require at least 2 pixels per resolution element, which doubles the size of the grating in equation (7).
[i] In practice, astronomical gratings often produce ~5-10% scattering wings leading to psf profiles that are rather more Lorentzian than Gaussian.

Commercial devices deliver around 64 to 128 output channels at most. A "hero" device achieving 256 output channels has been achieved in a research lab (Delage et al 2004). These are sufficiently well separated at the focal plane to suppress cross talk. Therefore, a continuously dispersed spectrum with 512 overlapping spectral channels seems already to be feasible. With sufficient funding and re-engineering, it seems entirely plausible that 1024 overlapping channels can be reached with better than -3 dB insertion loss. An alternative approach is to feed the output of a low-order AWG to several tandem PEGs to achieve the same spectral coverage, although inevitably this will lower the overall efficiency by ~10% or more.

**Remaining issues.** Astronomical fibres are typically multimoded, although future spectroscopic instruments are likely to operate with few-moded fibres (Horton & Bland-Hawthorn 2006). A large core fibre allows far more light to be inserted into the slab waveguide compared to conventional single-mode use.

A good discussion of multimode PEGs is give by McMullin et al (2002) where they consider a low-order glass device with 22 excited modes. The change of direction on reflection of the diffracted rays induced by modal dispersion is

$$\delta\beta \approx -2 \tan \beta \, (\delta n_a/n_a)$$

Once again, this shift is only likely to be negligible in low order devices. A moderately multimoded PEG has another advantage. In theory, optical modes in an optimized slab waveguide should propagate without loss, but in practice waveguide imperfections (e.g. facet roughness) can cause mode conversion. That the light distribution within a PEG is modally scrambled is evident from Figs. 3 and 6 in McMullin et al (2002). It follows that a moderately multimode PEG is likely to be more efficient as long as the modal dispersion produces negligible cross talk at the focal plane.

## 5. DISCUSSION

There are several factors that argue in favour of miniaturizing spectrographs in the years ahead, one of which was given in §1. At or near the diffraction limit, the focused spot is independent of the telescope aperture, and has a size that is compatible with photonic devices. If the light can be efficiently coupled into the device, it can remain within the device and be manipulated by it, before being imaged at the detector. The light does not need to see an air-glass boundary again. The cross-talk performance of telecomm devices already indicates that light scatter, birefringence and polarization effects can be managed to a high degree. Instruments based on integrated circuit technology will be more easily scaled to larger sizes, cheaper to mass produce, easier to control, and much less susceptible to vibration and flexure. Conventional optical design is now replaced by photonic engineering supported by a billion dollar R&D investment.

In recent years, there have been serious advances in achieving efficient suppression of the OH night sky emission using fibre Bragg gratings embedded within optical fibres (Bland-Hawthorn et al 2004). The night sky problem is widely viewed as one of the fundamental limitations to carrying out observational cosmology on large telescopes in the decades ahead. In order to exploit this technology, light needs to be concentrated into optical fibres. If we are to go to the trouble of concentrating light into small fibre cores, it makes sense to use these fibres as pre-filters to R~1000 photonic echelle gratings. We see this as a very exciting synergy between two new technologies.

Micro-spectrographs will have other serious advantages. Wide-field surveys continue to dominate astronomy and astrophysics, and will do so for decades to come. With the advent of AO correction, "wide field" takes on new meaning. We can talk about an AO-corrected, wide field if the sampled format involves thousands or tens of thousands of pixels on a side, regardless of the size of the sky field (Bland-Hawthorn 2006). Thus, wide field positioning systems will continue to dominate astronomy for any level of AO correction.

We can now envisage autonomous robotic positioners ("starbugs"; McGrath & Haynes 2006) carrying a stack of micro-spectrographs fed by a microlens array that sees the sky at the top of the stack. The data need not be transported by an optical relay system away from the focal plane: the stack of PEGs can be bonded directly to a custom-made infrared array or linear detector controlled by an ASIC. The stored data can either be transported away by optical fibre or, if there are to be no wires attached to the bugs, by individual "Bluetooth" or WiFi connections that have comparable footprints to the PEGs. Power can be supplied either by rechargeable batteries, or by grid lines in the base plate that defines the focal plane. This would allow the bugs to rove without tripping over wires, and would bypass difficulties associated with accurate alignment and metrology in optical relay systems.

Arguably, the greatest prospect of producing a viable "spectrograph on a chip" would be to realize a million element IFU – the so-called MEIFU concept – first envisaged by Content, Morris & Dubbeldam (2003). These already exist on target planes within particle accelerators that make use of bundles with $10^6$ or more optical fibres. We envisage a million "few mode" or "single mode" fibres feeding a thousand circuit boards, each with a thousand "integrated photonic spectrographs" illuminating linear IR detectors integrated onto the boards.

We have begun to address integrated photonic spectrographs and plan to demonstrate these within the lab in the next few years. There are numerous issues to resolve. We seek to achieve ~1000 contiguous spectral channels in the near infrared with losses of –3 dB or better. This performance is close to being achieved in tandem PAWGs but these are "hero" lab devices filling large wafers for demonstration purposes. These devices are largely overengineered – the output channels define a Shah function (i.e. picket fence with gaps in between) where the light is coupled into ~1000 individual waveguides. These devices are not ideally suited to astronomy without substantial investigation and development, except in one important respect (discussed in §3). There is another respect in which these devices might prove to be important: the ability to achieve excellent cross-talk separation between channels. We can envisage schemes in which unwanted spectral channels (e.g. OH night sky lines) are coupled into waveguides and summarily rejected.

Photonic echelle gratings are much closer to the needs of the astronomy community, and our first devices will be based on this technology. Here, again, there are numerous issues to resolve. We seek to produce devices that can operate with "few mode" input fibres to ensure higher coupling efficiencies than is possible with "single mode" fibres (Horton & Bland-Hawthorn 2006). Secondly, the insertion losses must be –3 dB or better, and the output response be as flat as possible in wavelength, and over as broad a spectral range as possible. PAWGs can be used to break up the spectral band into sections that are treated by individual PEGs (tandem arrangement). Integrated photonic spectrographs may seem fanciful at first glance, but we believe this is a natural development for astronomy in the mid to long term. The development work is supported by a billion dollar R&D investment to date fuelled largely by the telecomm industry. It is our view that a relatively modest investment will be needed to produce the first useful devices.

**ACKNOWLEDGMENTS.** We are indebted to David Moss and Andre Delage for their insights on integrated photonic circuits.